\documentclass{emulateapj}
\usepackage{amsmath}
\usepackage{amssymb}

\newcommand{\etal}{{\it et al.}}
\newcommand{\arcm}{{$^\prime\,$}}
\newcommand{\arcs}{{$^{\prime\prime}\,$}}    
\newcommand{\z}{{z}}    

\begin{document}

\title{Ubercalibration of the Deep Lens Survey}


\author{D. Wittman, R. Ryan and P. Thorman\\
Physics Department, University of California, Davis,  CA
95616\\ dwittman@physics.ucdavis.edu}



\begin{abstract} 
  We describe the internal photometric calibration of the Deep Lens
  Survey, which consists of five widely separated fields observed by
  two different observatories.  Adopting the global linear
  least-squares (``ubercal'') approach developed for the Sloan Digital
  Sky Survey (SDSS), we derive flatfield corrections for all observing
  runs, which indicate that the original sky flats were nonuniform by
  up to 0.13 mag peak to valley in $\z$ band, and by up to half
  that amount in {\it BVR}.  We show that application of these
  corrections reduces spatial nonuniformities in corrected exposures
  to the 0.01-0.02 mag level.  We conclude with some lessons learned
  in applying ubercal to a survey structured very differently from
  SDSS, with isolated fields, multiple observatories, and
  shift-and-stare rather than drift-scan imaging.  Although the size
  of the error caused by using sky or dome flats is instrument- and
  wavelength-dependent, users of wide-field cameras should not assume
  that it is small.  Pipeline developers should facilitate routine
  application of this procedure, and surveys should include it in
  their plans from the outset.
\end{abstract} 

\maketitle


\section{Introduction}

As CCD camera fields of view have become dramatically larger in
the past decade, ensuring uniform photometry has become more
challenging.  Traditionally, it is assumed that the camera can be
uniformly illuminated, by sky emission or by an illuminated screen,
and dividing by the response of each pixel to this illumination is all
that is necessary to make images internally consistent.  But the
sensor may not be uniformly illuminated due to vignetting or to
nonuniform sources of illumination such as scattered light, moonlight,
or ghosting.


These problems are exacerbated when the field of view is large.  For
small fields of view, a reasonable workaround for nonuniform
illumination is to observe a given star with a grid of pointings, so
that spatially-varying photometry can be directly identified and
modeled out.  This is impractical for very large cameras, where the
observation and readout times would be prohibitive.  However,
Padhmanabhan \etal\ (2008, hereafter P08) presented a
nearly-equivalent solution involving multiple stars.  In any set of
overlapping exposures or drift scans, some stars are multiply
observed, in different parts of the camera, and these observations can
be used to constrain a model of the spatial variation of the camera
response.  This technique, widely referred to as ubercal, analyzes the
survey data per se, and does not require additional calibration data.
This neatly separates the problems of internal and external
calibration.

As more and bigger imaging surveys come online, ubercal is sure to be
an important tool.  P08 applied it to a drift-scan survey, the Sloan
Digital Sky Survey (SDSS).  We apply it here to the Deep Lens Survey
(DLS) which, like the vast majority of imaging surveys, is a
shift-and-stare survey.  The DLS has some other peculiarities, such as
widely separated fields and multiple observatories, that make it an
informative example for the application of this technique. In
\S\ref{sec-dls} we describe the DLS and the corresponding structure of
the ubercal model; in \S\ref{sec-fit} we describe the fitting process;
in \S\ref{sec-res} we show the results and examine their robustness;
in \S\ref{sec-impact} we report the impact on survey photometry
accuracy; and we present conclusions in \S\ref{sec-con}.

\section{Ubercal formalism for the DLS}\label{sec-dls}

The basic idea of ubercal is to write down a model describing the
observing process (true magnitudes are offset by a zeropoint, an
airmass term, and a ``flatfield'' term describing the spatial
variation of the camera response), and constrain the model parameters
using the measured instrumental magnitudes.  Before writing a specific
model for the DLS, we first describe the structure of the DLS.

\subsection{Structure of the DLS}

The DLS consists of five well-separated 2$^\circ \times 2^\circ$
fields (Table~\ref{tab-coords}). The northern fields (F1 and F2) were
observed using the Kitt Peak Mayall 4-m telescope and Mosaic
prime-focus imager (Muller \etal\ 1998).  The southern fields (F3
through F5) were observed with a similar setup at the Cerro Tololo
Blanco 4-m telescope.  Each Mosaic imager consists of a 4$\times$2
array of three-edge-buttable 2k$\times$4k CCDs, providing a
35\arcm\ field of view with 0.26\arcs\ pixels and minimal gaps between
the devices.  Each DLS field is divided into a 3$\times$3 grid of
40\arcm$\times$ 40\arcm\ subfields.  These subfields are slightly
larger than the Mosaic field of view, but are synthesised with dithers
of up to 800 pixels (208\arcs).  In some cases, dithers were much
larger.  For example, CCD 3 was dead during two Cerro Tololo runs, and
we dithered by an entire chip width (8.5\arcm).  In a small number of
exposures, telescope pointing errors resulted in very large dithers
and/or odd pointing centers, further tying the subfields together.

\begin{table}
\caption{DLS field centers (J2000) and mean extinctions from Schlegel, Finkbeiner \& Davis (1998).}
\label{tab-coords}
\begin{tabular}{|c|c|c|c|c|}
\hline
Field & RA & DEC & l,b & E(B-V) \\ \hline
F1 & 00:53:25.3  &    +12:33:55   &    125,-50 & 0.06 \\
F2 & 09:18:00    &    +30:00:00   &    197, 44 & 0.02 \\
F3 & 05:20:00    &    -49:00:00   &    255,-35 & 0.02 \\
F4 & 10:52:00    &    -05:00:00   &    257,47  & 0.025 \\
F5 & 13:55:00    &    -10:00:00   &    328,49  & 0.05 \\ \hline
\end{tabular}
\end{table}

The DLS was observed in four filters, {\it BVR}$\z$.  $R$ was observed
when the seeing FWHM was $<0.9$\arcs; if the seeing was worse, the
choice of filters was dictated by moon phase, airmass, and past
progress in other filters. Thus, most nights contain images in
multiple filters, but few nights contain images in all four filters.
There were 25 observing runs of 2-6 nights each.  On a few runs,
southern fields (F4 at declination -5$^\circ$ and F5 at -11$^\circ$)
were observed from Kitt Peak. A northern field (F1 at declination
+12$^\circ$) was observed on only one Cerro Tololo run, on a
nonphotometric night.  At both sites, equatorial standard star fields
(Landolt 1992, Smith \etal\ 2002) were observed on photometric nights.
More details about the DLS survey design and image processing can be
found in Wittman \etal\ (2002) and Wittman \etal\ (2006) respectively.
A paper describing the final data release details (Wittman \etal, in
preparation) is forthcoming.

\subsection{Photometric model}

We adopt the following model for the instrumental magnitudes measured
through a given filter:
\begin{equation}
m_{instr} = m_{true}^{(s)} + z^{(r)} + k^{(n)}A +
f^{(r,c)}(x,y) \label{eqn-model}
\end{equation}
where $z$ is a run-dependent zeropoint, $A$ is the
airmass of the observation, $k$ is the extinction per unit airmass,
$f$ is the flatfield function, and superscripts indicate indexing by
star $(s)$, run $(r)$, night $(n)$, and CCD $(c)$ identification numbers.
Filter superscripts are not explicitly written here because the
fitting procedure is carried out totally independently in each filter.

We choose $f$ to be a second-order polynomial in the CCD coordinates
$x$ and $y$, with an $xy$ cross-term making a total of 6 parameters
per CCD.  There is no requirement that the flatfield be continuous
across CCD boundaries.  The primary motivation for allowing
discontinuities is that sky flats have already been applied to the
data, so we are really modeling systematic errors in the sky flat
process, which may include discontinuous jumps at the CCD boundaries.
The flatfield magnitude offset at the center of the first CCD is fixed
at zero to remove the degeneracy with overall throughput parameters
such as the zeropoint. This leaves 47 flatfield parameters to
constrain for each run (eight CCDs times six polynomial parameters,
minus one fixed offset).  Each run has its own flatfield model,
because sky flats were derived and applied on a per-run basis.


Each run also has its own zeropoint, so that the model can track changes in
non-atmosphere-related observatory throughput such as might occur when
the telescope optics are recoated.  Changes in CCD gain would also be
absorbed into this factor.

Airmass terms are, in principle, fit for each night.  In practice,
many nights were not photometric, and thus cannot be fit this way.  On
these nights, each image receives its own airmass term.  This
conceptually the same as receiving its own zeropoint, but happens to
be simpler to implement in our code.  A few images were taken with
nonstandard exposure times.  For fitting purposes, we converted the
instrumental magnitudes from these exposures to the instrumental
magnitudes that would have been observed had the standard exposure
time been used.

We choose not to model any time variation of the airmass term. P08
suggest a time term of 1 mmag per unit airmass per hour, based on data
from the SDSS photometric calibration telescope.  The sign is such
that the atmosphere grows more clear as the night goes on.  We have no
information on similar trends at Cerro Tololo and Kitt Peak.  DLS
differs from SDSS in being able to use only one filter at a time, so
that time baselines in a given filter tend to be much shorter.  Over a
three-hour time baseline, a change of 1 mmag per unit airmass per hour
is quite small.  Furthermore, because DLS has fixed fields rather than
a drift-scan strategy, airmass is often a monotonic function of time
in a given filter on a given night, in which case much of the time
variation (if it even exists) can be absorbed into the airmass term.
Therefore, we do not include time variation as part of the model.
This is further justified by the general absence of time trends in the
residuals.

\subsection{Spatiotemporal degeneracy}\label{subsec-spatiotemporal}

Our primary motivation for this work was to derive the flatfield
corrections, because the DLS photometry had been showing spatial
dependence.  Recall that each subfield was imaged 20 times in each
filter, with $\sim 5$\arcm dithers providing some overlap between
adjacent subfields.  Intuitively, one might think that this large
number of dithers would be sufficient to constrain the flatfield
corrections by itself.  However, there is an astonishing degeneracy
between spatial and temporal variations, as follows.

Consider an imaging device with a sensitivity variation $f(x,y) = ax +
by$ (in magnitudes here for convenience) and a time-varying zeropoint
$z(t)$.  For now we will not think of $z(t)$ as a function, but just
as a collection of zeropoints at the times $t_i$ when various
exposures were taken.  So a star at position $x,y$ will have
instrumental magnitude
\begin{equation}
m_{i} = m_{true} + ax + by + z(t_i)\label{eqn-2}
\end{equation}
Now consider a set of exposures with dither offsets $x_i,y_i$.  The
difference between instrumental magnitudes of a given star on the
$i$th and $j$th exposures is
\begin{equation}
m_{i} -m_{j} = a(x_i-x_j) + b(y_i-y_j) + z(t_i)-z(t_j)\label{eqn-3}
\end{equation}
This quantity is identical for all stars and galaxies, because
they all experience the same spatial shift.

Now consider an alternative model where the flatfield slopes are quite
different:
\begin{equation}
m_{i} = m_{true} + (a+\lambda_1)x + (b+\lambda_2)y + z(t_i)
-\lambda_1 x_i - \lambda_2 y_i 
\end{equation}
with $\lambda_1$ and $\lambda_2$ being arbitrary numbers making it
clear that this model is different from the previous model.  Recall
that x and y are the star's position on the detector, while $x_j$ and
$y_j$ are the dither shifts, so the last two terms affect the entire
exposure, effectively changing its zeropoint.  Now the difference
between exposures is
\begin{align*}
m_{i} -m_{j} &= (a+\lambda_1)(x_i-x_j) + (b+\lambda_2)(y_i-y_j) \\
& + z(t_i)-z(t_j) - \lambda_1(x_i-x_j) - \lambda_2(y_i-y_j)\\
&= a(x_i-x_j) + b(y_i-y_j) + z(t_i)-z(t_j)\tag{5}\label{eqn-5}
\end{align*}

Equations~\ref{eqn-3} and \ref{eqn-5} are the same, demonstrating that the
observed magnitude difference is identical in these two cases, despite
their very different implications for the flatness of the detector
sensitivity, and therefore for all the photometry derived therefrom.
This is a degeneracy.

Note that the degeneracy parameters $\lambda_1$ and $\lambda_2$ do not
depend on the position of the star on the detector, only on the dither
offset $x_i-x_j,y_i-y_j$.  Therefore, the density and spatial
distribution of stars is irrelevant.  Furthermore, the $\lambda_i$
depend only on the $i$th exposure's dither offset, not on any
properties of the other exposures.  Therefore, the number and spatial
distribution of dithers is irrelevant.  This is a true degeneracy as
long as the zeropoints have freedom to soak up any effect of the
flatfield slope.  But with observations in photometric conditions, we
can put a strong prior on $z(t_i)-z(t_j)$, thus constraining the
linear flatfield terms $a$ and $b$ to similar precision in units of
the dither size.  Surveys with substantial fractions of nonphotometric
time should be aware of this issue.  For example, they should ensure
that the nonphotometric time is not overly concentrated in any one
filter, lest it become impossible to solve for the flatfield
corrections in that filter.

This degeneracy is grasped much more intuitively and directly by
considering a technique used for small cameras, dithering a given star
around the field of view to directly measure the flatness of the
photometry.  Clearly, that procedure would work only if done quickly
enough to eliminate the possibility of changes in throughput due to
anything other than star position.  Similarly, ubercal requires a
strong temporal or airmass-based constraint on each exposure's
throughput.  

One might think that another way to fix the degeneracy would be to
rotate the CCD.  Although this does constrain the gradient, there are
still rotational modes which are degenerate with the time behavior.
We have confirmed this with simulations.

If $f(x,y)$ contains higher-order polynomial terms, these are {\it
  not} degenerate.  For example, the signature of a 2nd-order term
might be stars on the left side of the CCD getting brighter and stars
on the right side getting fainter after a small shift to the left, and
vice versa for a small shift to the right.  This cannot be mimicked by
a zeropoint change.  However, the linear terms are independent of the
higher-order terms and thus will remain unconstrained even in the
presence of the higher-order terms.

\subsection{Other degeneracies}

There are other degeneracies modelers should be aware of.  Most of
these were recognised by P08, but are worth mentioning here in the
context of a very differently structured survey.

{\it Absolute calibration degeneracy.} The bulk of the fit parameters
are the ``true'' magnitudes of the stars.  Of course, these are not
true magnitudes in an absolutely calibrated sense.  There is a
degeneracy in which uniform increase in true magnitudes can be
compensated by a uniform decrease in run zeropoints.  Fixing this
degeneracy is precisely the problem of external calibration, which we
do not address here.

{\it Relative calibration degeneracy for isolated fields.} The true
magnitudes in one field are fixed relative to the true magnitudes in
another field only if both fields were observed from the same
observatory on the same photometric night.  Non-photometric
cross-observations do not help because per-image zeropoints are free
to absorb any offset in the true magnitudes of one field relative to
the other field.  Almost every photometric night of DLS observations
touched multiple fields, so most fields are well tied together.
However, there is a natural north-south split with few photometric
cross-observations.  The B, V and $\z$ observations contain at least one
photometric night with cross-hemisphere observations.  However, in R
there are none.  This could be fixed with external calibrations from
the standard star fields.  However, in the spirit of ubercal, we make
use of the fact that both observatories often observed the {\it
  same} equatorial standard star fields, in particular SA 98 and SA
101 (Landolt 1992).  Therefore, we included SA 98 and SA 101 data in
the ubercal fit, not to provide absolute calibration, but to
strengthen the ties between north and south.  (One could make ubercal
perform {\it both} of these functions by fixing $m_{true}$ for the
Landolt (1992) stars rather than considering them nuisance parameters.
However, this would require all ubercal photometry to be measured
within a large (14\arcs\ diameter) aperture to be consistent with
Landolt (1992).  Therefore, the absolute calibration is better done
outside the ubercal framework.)

{\it Airmass/time degeneracy.} On many DLS nights, there is a
monotonic relationship between airmass and time for some of the
filters, because a given filter was used for only a few hours before
or after a field transited.  Thus, a time-dependent term would be
degenerate with an airmass term.  This degeneracy does not prevent one
from producing internally consistent photometry, but it does severely
hamper physical interpretation of the fit parameters.  Consider a
time-dependent airmass coefficient of the form $k + k_1t$.  With $k_1$
fixed at zero, the size of $k$ is an indicator of the clarity of the
night, and very large or very small $k$ indicates some problem with
the data (most likely clouds if $k$ is very large, for example).  With
$k_1$ free, we find that the data will be slightly better fit, but the
time/airmass degeneracy will probably drive $k$ and $k_1$ to
nonphysical values.  We therefore choose not to fit for $k_1$.

\section{Fitting}\label{sec-fit}

The procedures described below were carried out independently in each
filter, unless otherwise noted.

\subsection{Input data}\label{sec-input}

We ran SExtractor (Bertin \& Arnouts 1996) to produce a catalogue for
each CCD of each exposure, and then matched the catalogues in RA and DEC
to produce a master catalogue for each widely separated field. (This is
an exception to the per-filter rule; {\it all} images in a given field
were matched to ensure consistent astrometry across filters.)  Because
the eventual goal is to stack the images onto a grid with uniform
pixel size, we added master $x,y$ columns corresponding to the desired
output grid.  We also assigned a unique identification number to each
object in the master catalogues.  We then matched each CCD catalogue to
the master catalogue and fit for polynomial coordinate transforms
between CCD coordinates and uniform-grid coordinates.  We corrected
the photometry of each CCD catalogue by the Jacobian of the coordinate
transform, to correct for the varying pixel size without resampling
the pixels.  See the DLS data release paper (Wittman \etal, in
preparation) for a more complete explanation of this correction.  As
noted above, in a few cases we also adjusted the magnitudes to the
standard exposure time.

We then cleaned the catalogues.  We rejected observations with
SExtractor FLAGS$>$0, and cut on object peak intensity (SExtractor's
FLUX\_MAX output parameter).  On this attribute, we imposed both a
maximum (20\% below saturation after taking into account the sky level
which is not included in FLUX\_MAX) and a filter-dependent minimum
designed to prevent the catalogues from becoming extremely large.  The
minimum was set around 35$\sigma$ (for the peak pixel, not the entire
detection) in $\z$ band, and higher in the other bands.  Roughly 100
objects per CCD pass these cuts, yielding roughly 800 objects per
exposure and roughly 800,000 objects per filter.  The time required to
fit a catalogue of this size turned out to be surprisingly short (see
\S\ref{subsec-fitting}), so that the catalogs could have been made
more inclusive.  However, some lower signal-to-noise ratio (S/N)
cutoff will always be necessary, as low-S/N objects add limited
constraining power while increasing the compute time linearly.  We
prefer a conservative S/N cut because SExtractor overstates the S/N
by not accounting for sky modeling errors.

We used both stars and galaxies.  P08 used only stars ``to avoid
subtleties of extended object photometry.''  However, these subtleties
are mostly due to the range of galaxy profiles and morphology.
Because we are only concerned with {\it relative} photometry of the
{\it same} galaxy appearing in different places on the sensor, galaxy
photometry should be usable for this purpose.  In tests, we found
SExtractor's MAG\_AUTO (the magnitude within an aperture of size and
shape determined by the object itself) to be relatively immune to
seeing variations, so we used MAG\_AUTO for the fit.  

However, the stated uncertainties on MAG\_AUTO are unrealistically
small ($\sim$ 1 mmag) because they are based on photon noise.  For
bright objects, where photon noise is relatively small, measurement
uncertainty is likely dominated by background modeling errors.  At the
bright end, we found a noise floor of $\sim$0.02 mag as determined by
the rms differences between magnitudes measured for the same object on
pairs of images closely matching in time, airmass, and detector
position.  This value is only a rough approximation, as it is
certainly filter-dependent, and may decrease after taking flatfield
corrections into account, but it is much more realistic than the
nominal uncertainties, so we assign an uncertainty of 0.02 mag to each
measurement.  When interpreting the $\chi^2$ values from the fit, we
must keep in mind the roughness of the approximation here.

The cleaned catalogue in any given filter contains 5-7$\times 10^5$
observations and 6-7$\times 10^4$ unique objects, depending on filter.
The cleaned catalogue does not contain any objects observed only
once. Thus it should not contain cosmic rays or asteroids, except
where they happened to land on top of real objects.  In
\S\ref{subsec-fitting} we describe how to reject those observations.


\subsection{Photometricity of nights}\label{sec-nights}

Before using ubercal, our previous method of determining photometric
offsets between exposures was by matching exposure catalogues against
a master catalogue.  These offsets, along with observing logs and
standard star observations taken on nights thought to be photometric
by the observers, gave us a good starting point for marking nights as
photometric (and thus requiring a single $k^{(n)}$ parameter for the
night) or nonphotometric (and thus requiring an exposure-specific
magnitude offset).  However, we found (as did P08) that ubercal
itself was the best tool for picking up on subtle levels of cloud.
Nonphotometric nights stood out as having many exposures with large
mean residuals from the fit.  When a night was identified as
nonphotometric, the $k^{(n)}$ parameter for that night was removed
from the model, one extinction parameter per exposure was inserted,
and the fitting was restarted from scratch.  The process was
iterative, with ever-better overall fits highlighting ever more subtle
photometricity variations.  As described below, we found that even the
best-fit nights have unmodeled variations, so that the cutoff between
photometric and nonphotometric is somewhat arbitrary.  We adopted a
rule calling a night photometric if the vast majority of its exposures
were modeled to within 0.02 mag.

\begin{figure}
\includegraphics[width=3.6in]{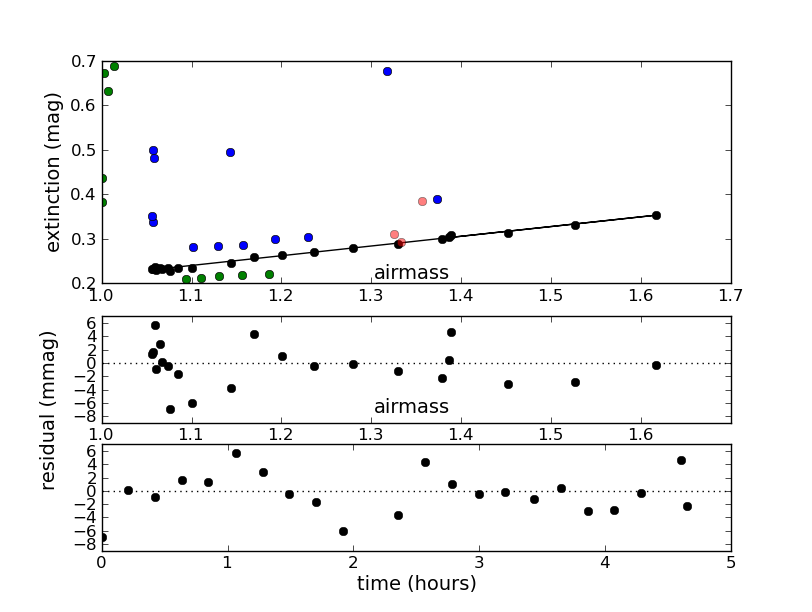}
\caption{Airmass fits and residuals for a run with one photometric and
  three nonphotometric nights in $B$ band.  Each point represents one
  exposure and each night is a different color.  The bottom panel
  shows the residuals from the fit (after averaging all objects in
  each exposure) as a function of time, and the middle panel shows the
  residuals as a function of airmass.  The top panel shows the model
  and data relative to the run zeropoint, such that the lines should
  intercept zero magnitudes at zero airmass.  
\label{fig-photometricitydemo}}
\end{figure}

Figure~\ref{fig-photometricitydemo} illustrates the process of
identifying nonphotometric nights.  Each color in the figure
represents a different night on one run (all in $B$ band).  We
rearrange Equation~\ref{eqn-model} to isolate the airmass part of the
model:
\begin{equation}
m_{instr}-(m_{true}^{(s)} + z^{(r)} + f^{(r,c)}(x,y)) = k^{(n)}A. \label{eqn-plot}
\end{equation}
We then average the left-hand side over the hundreds of objects
photometered in that exposure to obtain a single point for that
exposure.  These are the points plotted in the top panel of
Figure~\ref{fig-photometricitydemo}.  For one night (shown in black),
the points cluster around a straight line. This is the airmass model
on the right-hand side of Equation~\ref{eqn-plot}, which is also drawn
on the figure.  For two other nights (shown in blue and green), there
is a huge variance in the upper panel.  It appears that the telescope
was pointed at relatively clear sky for part of each night (points
lined up near the black line), but obviously cloudy sky for a
substantial number of exposures (points well above 0.3 mag on the
vertical axis).  The conclusion is that those nights were not
photometric, and therefore no airmass fit is drawn.  Instead, each
nonphotometric exposure is assigned a unique value of $kA$ in
Equation~\ref{eqn-plot}, equal to the mean of the left-hand side of
the equation.  For the night which was fit with a single parameter,
the residuals (as a function of airmass and of time) are shown in the
lower two panels.

The night shown in red has only three exposures in this filter, so the
evidence is not as strong as for the other nights, but variable cirrus
may have been a factor.  In cases like these, the presumption should
be nonphotometric.  The time and airmass baselines are short, so even
if the data were photometric they would not provide strong additional
constraints beyond their base nonphotometric value.  We stress that
even nonphotometric data help constrain the flatfields.  The
extinction due to clouds is unlikely to have spatial structure on the
scale of the camera (35\arcm), especially integrated over an exposure
time of 10-15 minutes.  Therefore, as long as other data in the survey
(on other runs, perhaps) constrain the true magnitudes of the objects
in the relevant areas of sky, the spatial variation of the photometry
in the nonphotometric exposure does provide information about the
flatfield.  So in the context of a larger survey with sufficient
photometric data, ubercal can solve for the flatfield even for a
completely nonphotometric run.

It is instructive to examine the residuals on photometric nights
(lower panels of Figure~\ref{fig-photometricitydemo}).  The fit is
good over a wide range of airmass and time.  The worst residuals are
6-7 mmag, whereas the measurement uncertainty on each residual is
about 1 mmag. The residuals are also structured in time, with a
quasiperiodic variation of $\sim 2$ hours and a peak-to-valley
amplitude of $\sim 10$ mmag.  These data do not tell us whether this
variation in extinction occurred in all directions, or only in the
particular field being followed.  It seems likely that a local
variation such as a band of very thin cirrus would move across the
field on faster timescales, and that on photometric nights the long
DLS exposures are more sensitive to global variations.  However,
surveys with much shorter exposure times will not necessarily find
spatial variations to be trivial, and may find a wider-field
boresighted cirrus monitor to be a good investment.

\begin{figure}
\centerline{\includegraphics[width=4in]{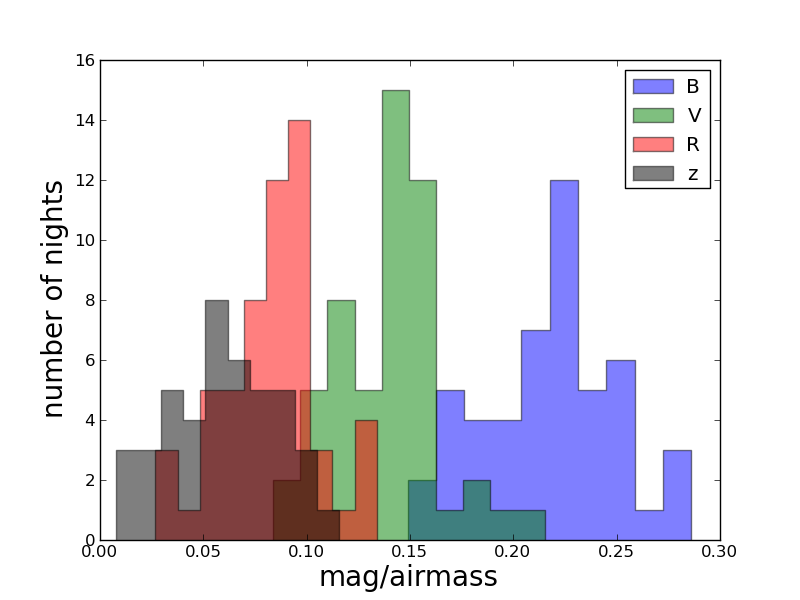}} 
\caption{Histograms of the best-fit nightly airmass terms, which
  cluster around the physically expected values in each filter.
  Nights which deviated greatly from these values were recognized as
  nonphotometric.  Typically, such nights were poor fits anyway.
\label{fig-amthist}}
\end{figure}

To be declared photometric, a night had to have not only a good
airmass fit, but a physically plausible airmass coefficient ($\sim$
0.05, 0.08, 0.14, and 0.22 in $\z$, $R$, $v$, and $B$ respectively).
Figure~\ref{fig-amthist} shows histograms of the best-fit nightly
airmass terms in the four filters, for those nights which were
considered photometric.  To prevent overfitting, the model enforces a
common zeropoint throughout a run. In other words, the fits for two
nights on the same run must intersect at zero airmass.  Although there
are physical mechanisms which could change the zeropoint at zero
airmass (recoating of the optics, for example), few of them are likely
to operate on a night-to-night timescale.  We could obtain fits with
slightly smaller residuals by freeing this constraint, but there is
little physical motivation for doing so.

Two of the nights in Figure~\ref{fig-photometricitydemo} are very
obviously cloudy, but many cases were more borderline, with $\sim 0.1$
mag departures from the best airmass fit.  Before definitively marking
a night as nonphotometric, we investigated other possible explanations
for a poor fit.  In particular, because good flatfield solutions were
our main goal, we investigated the possibility that a poor nightly fit
could be caused by poorly fitting flatfields rather than
nonphotometricity.  We did this by checking for correlations between
dither directions and residuals.  For example, with respect to the
first exposure of the five-exposure dither pattern, subsequent
exposures were shifted 3.5\arcm\ shift to the northeast, 3.5\arcm\
shift to the southwest, 1.75\arcm\ shift to the northwest, and finally
1.75\arcm\ shift to the southeast.  Thus, if a series of images fit
poorly due to poor flatfield solutions, one would expect the residuals
to reverse sign between the second and third, and/or the fourth and
fifth, exposures of a dither pattern, and for the residuals to be
largest on the second and third exposures.  In fact, we never found
this pattern, instead consistently finding the $\sim$2 hour
quasiperiodic variation described above.

In cases of borderline photometricity we also looked at behavior in
all filters on the same night.  If the collective behavior seemed
nonphotometric, we marked the night as nonphotometric.  Once marked as
nonphotometric for any reason, a night is considered nonphotometric in
all filters.

A few images were flagged as outliers despite being on apparently
photometric nights.  About half of these had clearly identifiable
causes, such as incorrect header information, or the last exposure of
a night pushing too far into dawn and corrupting the input photometry.
In the case of bad header information on a photometric night, the
image was returned to the photometric list if the correct metadata
(primarily airmass) could be reconstructed from observing logs.
Otherwise, the image was considered as a singleton, receiving its own
exactly fitted value of $kA$ just as if it were nonphotometric.  The
most interesting case was an exposure where apparently the guider
jumped stars at some point, resulting in double images of everything
(but one image several times fainter than the other). The fact that
this was not flagged manually at the time of observation demonstrates
the porosity of manual checks and the benefits of fitting all the
survey data simultaneously.

\subsection{Writing and running the code}\label{subsec-fitting}

The core fitting code was kindly provided by N. Padhmanabhan, written
in C++ with a conjugate gradient optimiser based loosely on the one in
Press \etal\ (1992).  This is not the same code used in P08; it was
written from scratch to assess the feasibility of ubercal for the
Large Synoptic Survey Telescope (LSST, Abell et al.\ 2009), and only
solved for per-image zeropoints.  We modified the code for DLS, adding
indexing by run, by night, and by CCD, and solving for polynomial
flatfield solutions, airmass terms, and per-run zeropoint offsets.  To
avoid potential numerical instabilities, the polynomial coordinates
are centered on each CCD and tracked in units of CCD widths rather
than pixels.  For convenience, the run zeropoints are referenced to
the center of CCD 1 rather than to an average over the entire array.

The code takes only 6 minutes to converge on a standard desktop
computer with $7\times 10^5$ data points and 6.5$\times 10^4$
parameters (mostly nuisance parameters, the true relative magnitudes
of the objects). We declare convergence when an iteration yields a
relative $\chi^2$ reduction of $10^{-9}$ or less.  Because the total
$\chi^2$ is about $\sim 6\times 10^5$ after convergence, this
criterion corresponds to a $\chi^2$ reduction of much less than one
per iteration, but due to the poorly defined photometric uncertainties
the value of $\chi^2$ is less important than the fact that further
iterations do not decrease it.  This level of convergence takes about
3000 iterations.  As an external check of convergence, we restart the
fitter with the best parameters but no memory of the gradient, which
forces the fitter to test a new direction in parameter space; the
result was always consistent with the original solution.

We checked for images with large ($\gtrsim 0.05$ mag) rms scatter in
the residuals, even if the mean residuals were acceptable.  A few
images with high sky levels due to approaching dawn were identified
this way.  These images appeared good to a casual inspection, e.g, the
sky was not very close to saturation.  Although they contributed
relatively little information to the survey due to their high sky
noise, the pre-ubercal judgment was to include them because they added
something.  However, the ubercal residuals were often highly
non-Gaussian, hinting at sky-modeling problems or other subtle
problems related to the bright sky.  Therefore, the judgment changed:
the small information contribution could be outweighed by other
problems, and we decided to exclude these images from the survey.

As mentioned above, asteroids and cosmic rays can land on top of real
objects and corrupt their photometry.  For these and other reasons, we
lightly clip the catalogue before fitting for the final solution. We
clip at 6$\sigma$ plus the mean residual of the worst-fitting exposure
(so that the worst-fitting exposure does not have many individual
objects clipped simply due to some overall characteristic of the
exposure).  The rms residual is about 0.02 magnitudes (ranging from
0.015 in R to 0.025 in $\z$), and the mean residual of the
worst-fitting exposure ranges from 0.014 in V to 0.025 in B, thus
setting a clip threshold of 0.10 to 0.17 magnitudes depending on the
filter.  Clipping removes the one outlying observation of an object,
rather than all observations of that object.  Manual inspection of
some of the clipped observations confirmed that the causes were
generally identifiable as cosmic-ray hits, proximity to an unmasked
bad pixel or bleed, foreground asteroid, etc.

After clipping, the reduced $\chi^2$ is about unity.  However, the
value of $\chi^2$ is not a rigorous indication of goodness-of-fit
because we manually set the instrumental magnitude uncertainties to
0.02 mag in the absence of realistic uncertainty estimates from
SExtractor.  The reduced $\chi^2$ ranges from 0.56 in $R$ to 1.65 in
$\z$ because the $R$ photometry is more precise than 0.02 mag ($R$
being the deepest filter and the one with the best seeing) and the
$\z$ photometry is less precise (due to the large sky noise).  Thus
the $\chi^2$ values indicate that the fits are plausibly good, but we
base our judgment of the fits mostly on other considerations such as
inspection of all the airmass plots and (as described below) physical
plausibility of the solutions, closed-loop correction tests, and
robustness against varying preparation of the input photometry.

\section{Results}\label{sec-res}

There are three main types of output: the nightly airmass terms, the
run zero points, and the spatial sensitivity functions (``flatfields''
for short).  We show some examples of each and offer physical
interpretations.

\subsection{Airmass Terms}\label{sec-amt}

We have discussed the airmass fits already, because they are crucial
to the iterative process of marking nights as nonphotometric and
converging on a final fit.  We therefore focus here on some 
examples which are atypical but which provide insight into the process.

\begin{figure}
\centerline{\includegraphics[width=3.6in]{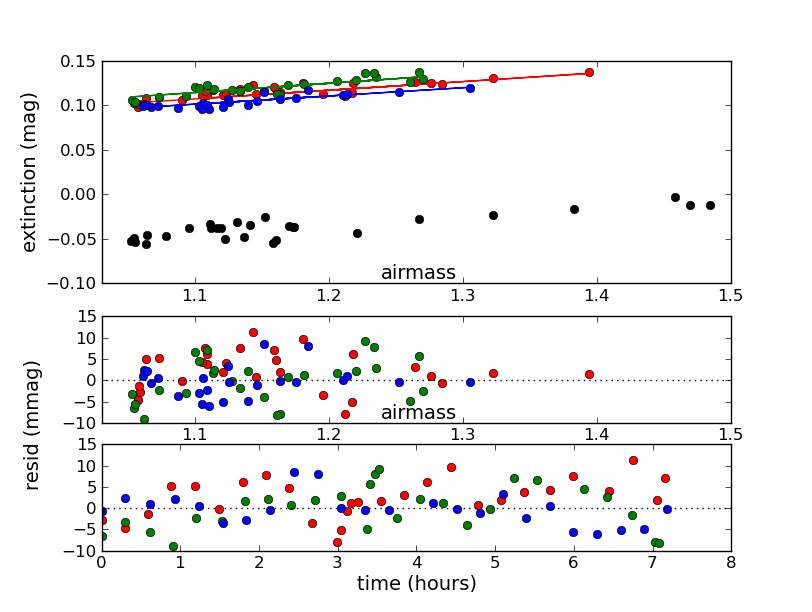}}
\caption{Airmass fits for the January 2001 CTIO run in R band.  Three
  nights (shown in green, red, and blue) were well fit by a common run
  zeropoint plus nightly airmass terms.  One night (shown in black)
  has a magnitude offset which is too large and too
  airmass-independent to be an atmospheric variation.  This kind of
  change could be caused by a change in overall scaling factor such as
  CCD gain.  \label{fig-amtweird}}
\end{figure}

Figure~\ref{fig-amtweird} shows the airmass fits for the January 2001
CTIO run in $R$ band.  Three nights (shown in green, red, and blue) were
well fit by a common run zeropoint plus nightly airmass terms.  In
other words, the fitted lines all converge to a point (the run
zeropoint) at zero airmass.  The first night (shown in black) has
similar airmass slope but is offset by about 0.15 mag, i.e. it could
never converge at the common run zeropoint.  Put simply, exposures
taken on this night contain about 15\% greater flux at any given
airmass.  Possible causes fall into several categories:
\begin{enumerate}
\item the camera collected 15\% more photons from each source on that
  night.  There is no plausible physical scenario enabling this.  If
  the other nights on the run were photometric with airmass terms
  $\sim$0.07, then even complete removal of the atmosphere would not
  suffice to increase throughput by this much.

\item the CCD amplifier gain differed by 15\% on that night.  This is
  unexpected, but not as implausible as the first scenario.
\item the data were changed after acquisition, for example, by
  multiplying all images from this night by 1.15, or applying a
  different flatfield than the other nights.  There is no evidence in
  the image headers for this.
\item the photometry procedure is not robust, for example being
  seeing-dependent.  However, the seeing on this night is not very
  different from the other nights on this run, and other runs with
  much larger seeing variations do not show this behavior.
\item exposure times were uniformly 15\% longer on this night.  This
  is implausible given the headers and observing logs.
\end{enumerate}

Only $R$ data were taken on this night, so behavior in the other
wavelengths, our cross-check of first resort, cannot be checked.  Two
other nights show similar behavior, but at a lower (4-5\%) level. One
of those nights contains observations in only one band, but the other
contains observations in three bands, and all three bands are
consistent.  Because different bands are reduced totally
independently, this makes it very unlikely that a data processing
misstep is responsible, and more likely that a gain change is
responsible.  Whatever the cause, the practical impact is that these
nights cannot be fit with the usual model of nightly airmass term and
overall run zeropoint.  In principle, they could be fit with a nightly
airmass term plus nightly zeropoint, but for convenience we simply fit
them with per-exposure zeropoints as if they were nonphotometric.

\begin{figure}
\centerline{\includegraphics[width=3.6in]{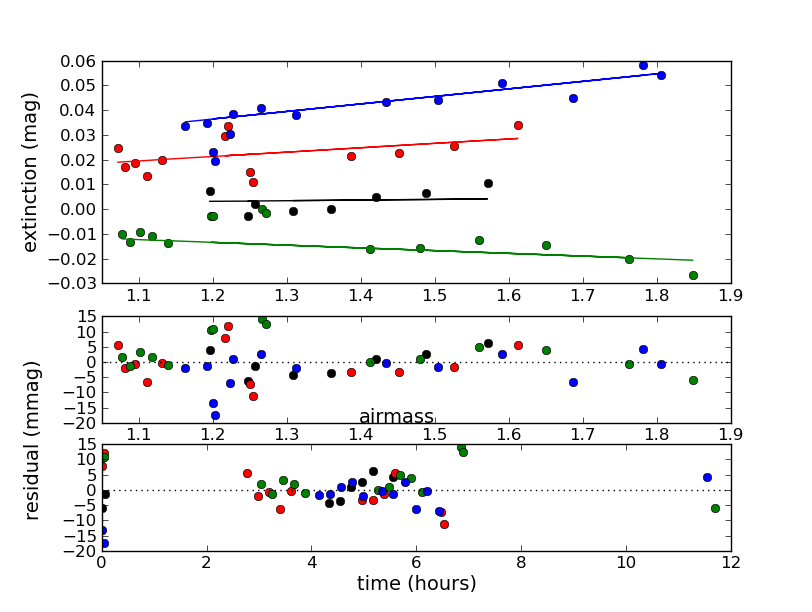}}
\caption{The best-fit airmass terms for the November 2000 Kitt Peak
  run are puzzling.  See text for details. \label{fig-negclouds}}
\end{figure}

In Figure~\ref{fig-negclouds}, we show a second example of unexpected
behavior of the airmass fits.  The night shown in green is best fit by
a negative airmass term.  This is completely unphysical, so the
natural conclusion is that the night had varying amounts of clouds
which happened to anticorrelate with the airmass.  However, if that
night had clouds, then the other nights on the run had even more
clouds, which also implausibly conspired to yield airmass fits with
small residuals.  There is no indication of this in the observing
logs, nor in the best-fit run zeropoint (which does not suggest
overall extinction greater than other runs), nor in the other filters
observed on these nights.  A second hypothesis might be that the fit
is forcing the night shown in green into unnatural behavior because
all the best-fit lines must converge at zero airmass.  But freeing
that night to float to its own zeropoint does not change the behavior
shown, nor does freeing each exposure change the pattern.  This effect
is seen only on the $\z$ band data from this run, so it could possibly
be related to the temperature dependence of the detector's $\z$ band
sensitivity.  These data would be fit by a scenario in which the
detector warms slightly with airmass (not with time) due to the
arrangement of (a presumably low level of) liquid nitrogen in the
dewar relative to the cold finger when pointing at our field.  

 We also explored some alternative hypotheses not involving
  umodeled physical effects.  We tested the local-minimum hypothesis
  by perturbing the solution and starting from a variety of places in
  parameter space; the fitter always returned to this solution with a
  few nights of negative airmass coefficients.  We also tested a
  hypothesis in which too much of the $\z$ data were nonphotometric,
  thus providing too few constraints to arrive at a physical solution.
  We tested this by fixing the magnitudes of objects appearing in the
  SDSS catalogue at their SDSS magnitudes.  This provided many
  hundreds of absolute reference points to better constrain the
  solution, but the negative airmass coefficients did not go away.  We
  conclude that an unmodeled physical effect is indeed at work, even
  if it is not the specific dewar-related effect described in the
  previous paragraph.  Fortunately, the best-fit parameters of
  interest (flatfield corrections and relative throughputs of each
  exposure) do not change substantially depending on how we model this
  run, so that we can adopt these parameters regardless of the
  physical interpretation.

\begin{figure}
\centerline{\includegraphics[width=3.6in]{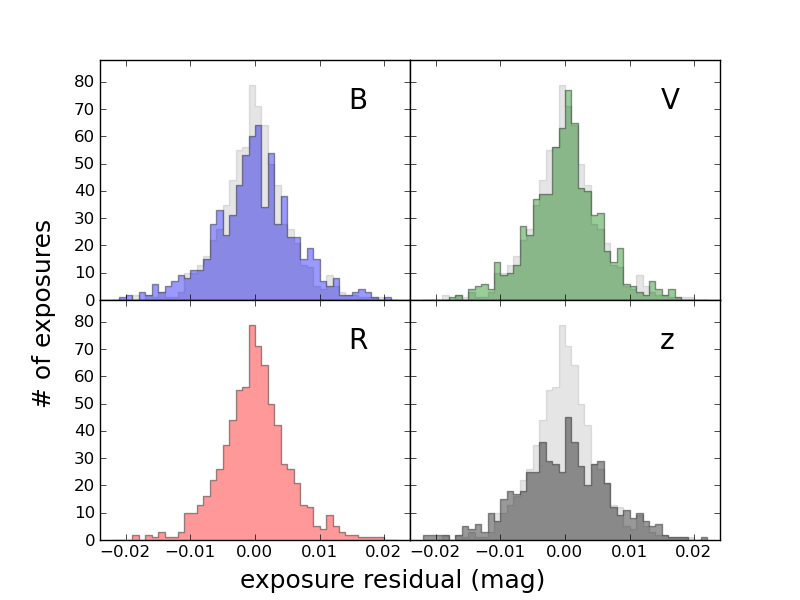}}
\caption{Histograms of the exposure residuals in each filter, for
  exposures in photometric conditions. The $R$ histogram is superposed
  on the $BVz$ histograms for comparison.  \label{fig-expresid}}
\end{figure}

We stress that these examples are interesting precisely because they
are atypical.  Most runs show no such unexpected behavior, yielding
fits similar to the uppermost three nights in
Figure~\ref{fig-amtweird} and residuals similar to the lower panels of
that figure.  The atmospheric variations are often modeled to within a
few millimagnitudes over a wide range of airmass.
Figure~\ref{fig-expresid} shows histograms of the exposure residuals
in each filter, for exposures in photometric conditions (recall that
the residual is identically zero for exposures in nonphotometric
conditions).  These histograms indicate that a simple airmass term is
sufficient to model many hundreds of photometric exposures to within
$\sim5$ mmag.

\subsection{Run Zeropoints}\label{sec-zpts}

\begin{figure}
\centerline{\includegraphics[width=3.6in]{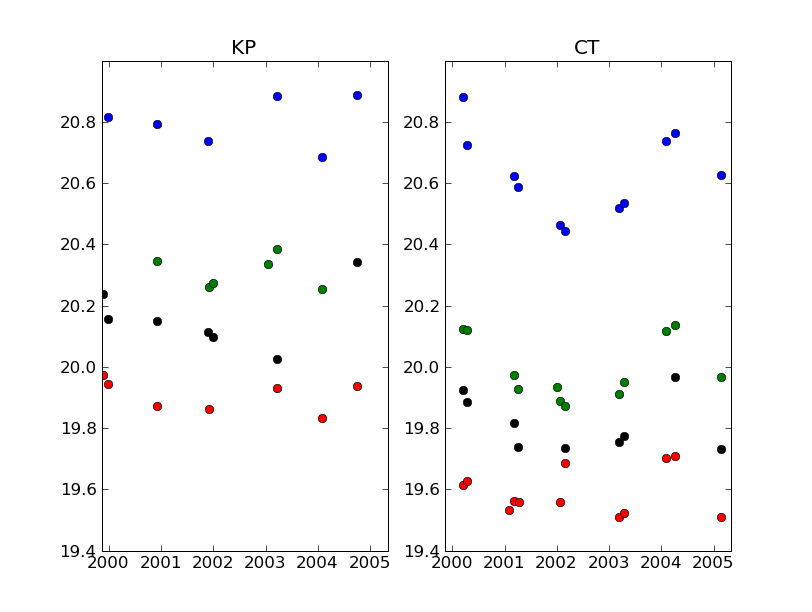}}
\caption{Run zeropoints, as defined in Equation~\ref{eqn-model}, for
  each run with at least one photometric night in a given filter ($B$,
  blue; $V$, green; $R$, red; $\z$, black) at Kitt Peak (left) and
  Cerro Tololo (right). \label{fig-zpts}}
\end{figure}

Figure~\ref{fig-zpts} shows the run zeropoints for each run with at
least one photometric night in a given filter.  Only {\it relative}
zeropoints are meaningful, and only within a given filter.  The most
striking feature is that the four filters vary so nearly in unison,
especially at CTIO, despite the completely independent fit in each
filter.  This, along with the month-to-month stability, suggests that
the variations reflect real long-term changes rather than noise.  (The
referencing of the zeropoint to the center of CCD 1 rather than to an
average over the entire array causes only $\lesssim$0.02 mag shifts in
this plot, which we neglect hereafter.)

Because the zeropoint appears on the right side of
Equation~\ref{eqn-model} rather than on the left, an increase in
zeropoint implies lower pixel values for a given star (this is
contrary to how observational astronomers usually think about a
zeropoint).  Upward excursions in zeropoint could be caused by
decreases in the atmosphere-independent part of the throughput, higher
CCD amplifier gain (i.e., lower pixel value per photoelectron in the
device), or similar factors. The variations do not appear to be
correlated with recoatings of the optics at CTIO (A. Walker, private
communication). We do have evidence of undocumented gain changes on
shorter timescales (\S\ref{sec-amt}), so longer-term drifts in gain
could be responsible for the coherent part of the time
variation. However, because the variation at shorter wavelengths is
larger than at longer wavelengths, this cannot be the entire
explanation. Wavelength-dependence suggests a physical cause, and the
blueness of the effect rules out some physical causes such as CCD
temperature changes. We conclude that there are long-term
($\sim$2-year) variations in hardware throughput due to as-yet
unexplained causes.

\subsection{Flatfield Corrections}\label{sec-fff}

Figure~\ref{fig-fff} shows the flatfield corrections in magnitudes,
averaged over all runs at Kitt Peak (left column) and nearly all runs
at Cerro Tololo (right column), in $B$, $V$, $R$, and $\z$ (top row to
bottom row). This is a visualisation of the last term in Equation 1,
so a larger positive number means that a star in the sky-flattened
images appears fainter than it would have if the sky flats had been
perfect. The variation from run to run is much smaller than the
systematic differences between telescopes and among filters, so we
begin by discussing those patterns:

\begin{figure}
\centerline{\includegraphics[width=3.6in]{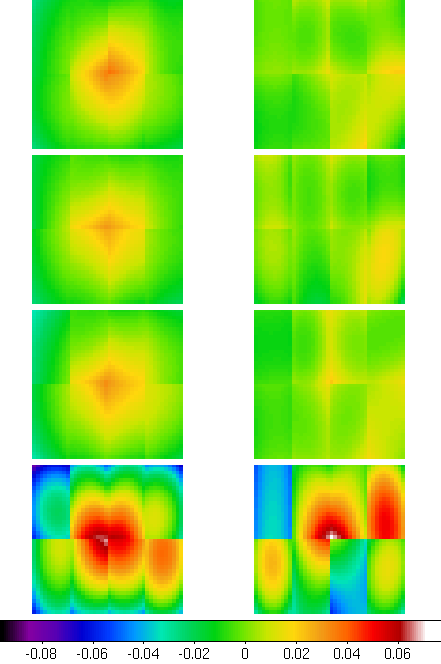}}
\caption{Flatfield corrections in magnitudes for Mosaic at Kitt Peak
  (left column) and Mosaic 2 at Cerro Tololo (right column), in $B$,
  $V$, $R$, and $\z$ (top row to bottom row). Each image is a mean
  over many observing runs; the variation from run to run is much
  smaller. See text for physical interpretation.
\label{fig-fff}}
\end{figure}

\begin{enumerate}
\item The Kitt Peak and Cerro Tololo corrections clearly form two
  different families. We stress that there is nothing in the model to
  enforce or even suggest that this should be the case. Each run has
  its own set of flatfield parameters, unconnected to the other
  runs. The fact that the flatfield parameters for the Cerro Tololo
  runs are similar to each other and dissimilar to those for the Kitt
  Peak runs is therefore due entirely to the data rather than the
  model. This gives us confidence that the solution is representing
  real systematic errors in the flatfielding.
\item In {\it BVR}, the corrections are more or less continuous across
  CCDs. Again, we stress that this is not required or encouraged by
  the model, and the fact that data prefer it suggests that
  the data are responding to physical effects which operate mostly on
  Mosaic-wide scales. However, $\z$ band is not very contiguous
  because it is very sensitive to CCD thickness, which is obviously
  not continuous at the boundaries.  Because there can be real
  discontinuities, we retain the chip-by-chip model despite the
  dominance of the larger-scale patterns.
\item At each observatory, $B$, $V$, $R$ are similar to each other and
  very different from $\z$ band, which has by far the biggest
  corrections.  Again, this was not suggested by the model, which is
  completely independent in each filter.
\end{enumerate}

We attribute the differences between $\z$ and {\it BVR} to physical
factors.  First, across the $\z$ band the CCD sensitivity is a strong
function of CCD thickness and wavelength.  The discontinuous nature of
the $\z$ band corrections is likely due to varying device thickness.
Second, the larger amplitude of the radial pattern in $\z$ may be due
to the effective wavelength of the $\z$ filter changing across the
field, because it is an interference filter.


Although the hardware setups at Kitt Peak and Cerro Tololo are quite
similar, they are not identical. The telescopes have different prime
focus correctors, and some differences in the raw images produced by
Mosaic and Mosaic 2 are evident upon casual inspection, depending on
wavelength. Thus it is not surprising to see telescope- and
wavelength-dependent systematics reflecting our inability to
completely remove these effects in the image processing. The fact that
the data, unguided by the model, produce the physically plausible
patterns listed above lends confidence to the results.

The sign of the central pattern seen in the Kitt Peak corrections can
be interpreted as follows.  If there are too many sky photons near the
center (perhaps traversing the optical system through paths which are
forbidden to object photons), then dividing the raw data by the sky
flats will make objects near the center appear to be too faint, which
is the sign of the pattern shown here.  This is about a 5\% or 0.05
mag effect from center to edge in {\it BVR}, and a bit more than twice
as large in $\z$.  In $\z$ band at Kitt Peak an additional
complication is the presence in raw images of a ghost image of the
pupil, covering much of the array and with an amplitude $\sim$3\% of
the sky. Mosaic $\z$ images are therefore reduced with an extra step,
the IRAF task {\it rmpupil} which attempts to model and remove this
additive effect which is roughly centered on the array.
Undersubtracting this feature will lead to the same sign error shown
here, centered on the array.  However, the ghost pupil hypothesis does
not explain why the corrections derived here are sizable even in the
bands where the ghost pupil is not evident.  Furthermore, the
amplitude of any ghost pupil image undersubtraction is likely to be
negligible on the scale shown here, as only a $\sim$3\% error would
result in the case of no subtraction at all.

The central pattern seen in {\it BVR} at Kitt Peak is about the same
amplitude as the Jacobian correction, which is also centrally
symmetric.  However, we rule out an incorrect Jacobian correction as
the cause based on two lines of reasoning.  First, the correction is
applied the same way for the Cerro Tololo, which shows no such central
pattern.  Second, if ubercal were compensating for a mistake in the
input photometry, then the corrected final images would contain the
negative of this mistake.  Analysis of such images
(\S\ref{subsec-closeloop} and \S\ref{sec-impact}) demonstrates that this
is not the case.

The usual application of sky flats is known to be wrong for another
reason (Stubbs \etal\ 2007).  Sky flats represent one integral over
wavelengths, and object fluxes represent a different integral over
wavelengths, but the proper correction would involve dividing the
integrands, not the integrals.  If this is the cause of flatfielding
errors, the ubercal is performing a correction for the mean colors of
the objects in the input photometry.  However, given the Mosaic-wide
scale and central symmetry of most of the corrections derived here, we
favor an optics-related explanation for the bulk of these corrections.

For a given filter and observatory, the flatfield corrections are
fairly stable over time.  At a given camera location, the standard
deviation across runs is typically 4-5 mmag, an order of magnitude
smaller than the corrections shown here.  This consistency might
suggest constraining the flatfield corrections even more strongly by
solving for a single pattern per observatory and wavelength. However,
runs closer in time seem to be somewhat more consistent with each
other, suggesting that some of the time variation is real. Therefore,
we continue to model each run separately.

\begin{figure}
\centerline{\includegraphics[width=3.5in]{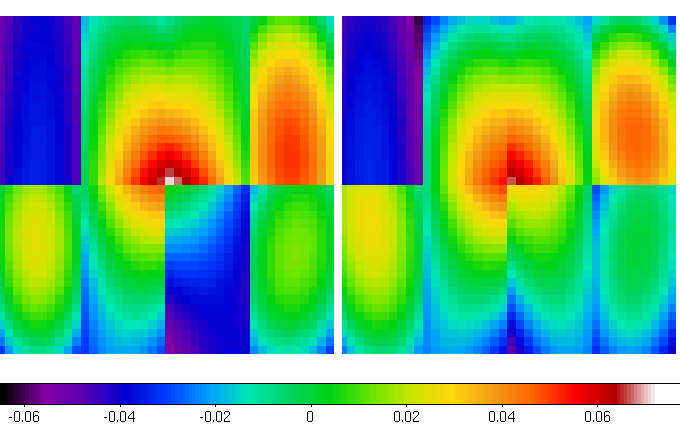}}
\caption{Flatfield corrections for Cerro Tolo in $\z$ before and after
  the replacement of CCD 3 (third from left in bottom row). As shown
  by the other CCDs, the flatfield corrections are generally stable
  over time although the model allows independent parameters for each
  run.  This one-time change of hardware clearly shows up in the
  ubercal-derived corrections.
\label{fig-ccdreplace}}
\end{figure}

One striking example of variation with time is when CCD 3 at Cerro
Tololo failed and was replaced.  Figure~\ref{fig-ccdreplace} shows the
$\z$ correction over the whole field before and after replacement.
CCD 3 has clearly changed, while the rest of the field has not changed
substantially.  Before replacement, the corrections for CCD 3 were
just as stable from run to run as the other CCDs shown here.  The {\it
  BVR} corrections for this CCD also changed at the same time, but are
of smaller amplitude.  Again, we stress that these patterns come from
the data because the model knows nothing about the CCD replacement.

\begin{figure}
\includegraphics[width=3.5in]{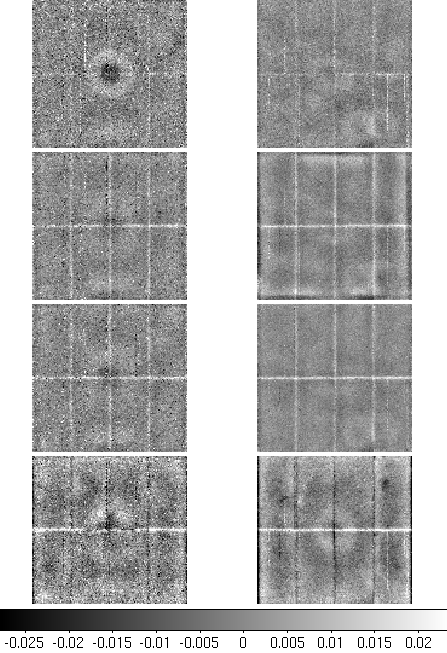}
\caption{Map of the residuals (in magnitudes) back onto the camera
  coordinates at Kitt Peak (left) and Cerro Tololo (right) for, from
  top to bottom, $B$, $V$, $R$, and $\z$.  For example, at Kitt Peak
  in $B$, objects in the center of the camera field are observed to be
  systematically brighter than modeled, revealing an inadequacy (at
  the 15 mmag level) of the second-order polynomial flatfield
  correction model.
\label{fig-spatialresids}}
\end{figure}

We can test the appropriateness of the polynomial model by mapping the
residuals back onto camera coordinates. To build up a statistically
meaningful map, we average over all runs in a given observatory/filter
combination. This is shown in Figure~\ref{fig-spatialresids} for
Mosaic (left) and Mosaic 2 (right) for the filters in the usual order.
At Kitt Peak, we see faint echoes of the ghost pupil, most clearly in
$B$, where it is barely visible in the raw images at the $<1\%$ level
but not modeled out in the image reduction stage.  At Cerro Tololo,
three ghost pupils are evident in most filters---a large one at the
center and two smaller ones at lower right---and some additional
process must be at work in $V$.  Despite the clarity of these
patterns, their amplitude is small, about an order of magnitude
smaller than the patterns captured by the polynomial model.  Their
spatial frequencies are also high, so that they will be washed out by
dithering.  A way to capture this high spatial frequency information
without adding multiple parameters would be to add a smoothed version
of the appropriate map back into the flatfield model as an extra term,
perhaps with a free parameter for its amplitude on each image or on
each run.

A few other features in these maps are worth noting.  Bad columns are
visible in places, indicating that the input photometry could have
been cleaned better.  More interesting is the edge behavior.  In {\it
  BVR}, objects near inside edges are consistently fainter than
expected (the white bands are not depictions of the CCD gaps), but
whatever causes this does not affect photometry near the outside
edges.  This inside/outside distinction points to a physical cause
rather than a SExtractor issue.  In $\z$ the sign of the effect
reverses for the long inside edges, and objects near the outside edges
are now affected as well.  These are small effects, but consistent
across observatories, so it is presumably related to the CCDs, readout
electronics, and/or image processing steps, which are largely the same
at both observatories.  Future surveys wishing to obtain very high
precision should examine spatial patterns in the residuals and use
them to refine their models.

\subsection{Closing the loop}\label{subsec-closeloop}

To implement the corrections, we modified our stacking software ({\it
  dlscombine}, Wittman \etal, in preparation) to perform the flatfield
corrections on the fly as it stacks.  We also configured it to use the
extinction corrections derived from ubercal, rather than from our
previous method of estimating extinction corrections.

\begin{figure}
\centerline{\includegraphics[width=3.6in]{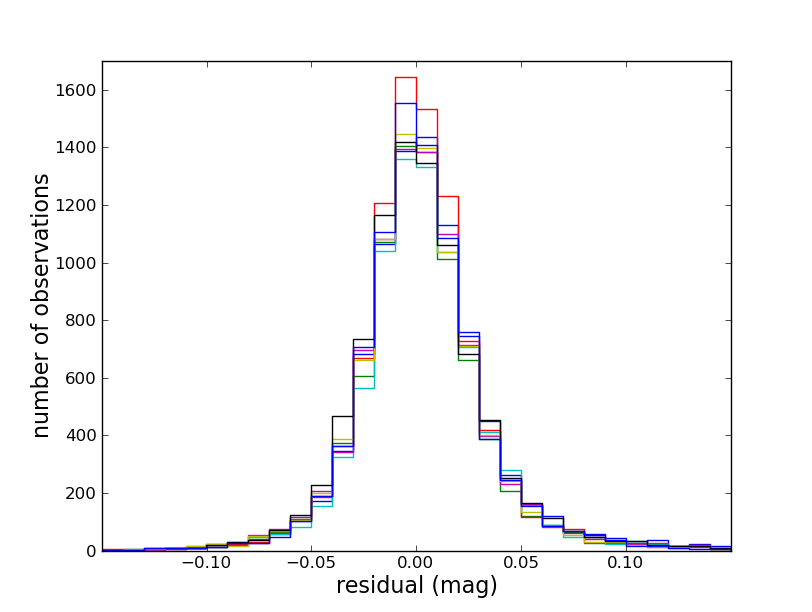}}
\caption{Histogram of residuals in the end-to-end test for field F2 in
  $\z$ band, with each CCD in the Mosaic represented by a different
  color.  Systematic offsets between CCDs, if any, must be very small.
\label{fig-residhist}}\end{figure}

We tested the corrections by having {\it dlscombine} write out
corrected postage-stamp images for each observation of each ubercal
object.  This is an end-to-end test involving many interrelated
pieces, such as applying the correct Jacobian determinant to the
ubercal input catalogues so that the combination of flatfield
corrections and repixelisation produces uniform photometry.  We
photometered the postage stamps using the same procedure as for the
ubercal input catalogues, and subtracted the model true magnitude for
each object.  The resulting residuals should have zero mean, with rms
similar to the rms residuals found in the ubercal fit.  More
importantly, any subset of residuals should have zero mean, so we can
subdivide the data by field, subfield, CCD, etc, and check for
systematics.  None of these attributes show any significant trends.
For example, Figure~\ref{fig-residhist} shows histograms of the $\z$
band residuals for the end-to-end test in field F2, subdivided by CCD.
There is no evidence for a systematic offset in any CCD.  A similar
comparison of fields yields no evidence for field-to-field offsets
either.  

The residual measured for any observation of any object with the
end-to-end test is very well predicted by the model residuals in
ubercal itself.  For example, if the ubercal fit left a +0.05 mag
residual for a particular observation of a given star, we find that
the measured magnitude of that star in that observation's corrected
postage-stamp image is indeed 0.05 mag fainter than the measured
magnitude of the same star in its other corrected postage-stamp
images.  The error is thus in the modeling, not in the application of
the corrections, although applying the corrections and
re-photometering does add some ($\sim$ 0.01-0.02 mag) extra scatter.
Thus, Figure~\ref{fig-residhist} is just a slightly broadened
histogram of the residuals estimated by ubercal itself.  The resulting
confidence in the model residuals is a key feature for rapid
refinement of the modeling process.  We can much more quickly search
the model residuals for patterns than we can apply the corrections and
search the corrected photometry for patterns.

The software used to produce the final DLS stacks is identical to that
used for the end-to-end tests, with only minor changes in the output
configuration to facilitate the postage-stamp analysis.  Thus the
procedure should (and in fact did; see \S\ref{sec-impact}) greatly
improve the uniformity of the photometry in the DLS stacks.
Nevertheless, two processes which do affect the final DLS photometry
are not tested by the methods of this section.  First is the actual
coaddition of pixels from different exposures.  The coaddition of
images with different point-spread functions (PSFs) introduces issues
which are explicitly not tested here.  Second, the method of producing
catalogues from the DLS stacks differs from the method used for
photometry here, because the prime consideration in constructing the
DLS catalogues is consistent measurement of {\it colors}, which involves
the PSFs of all the bands, whereas ubercal deals with only one at a
time.  These issues are discussed in more detail in the DLS data
release paper (Wittman \etal, in preparation).  Despite these details,
ubercal was by far the most important step for making the DLS
photometry uniform.

\subsection{Robustness}\label{sec-rob}

The results presented above are robust against marginal changes in the
model.  For example, for those nights which are borderline
nonphotometric, the flatfield corrections do not change substantially
when the status of the night is changed from photometric to
nonphotometric or vice versa.  Another type of model change
  discussed above was the use of SDSS photometry to fix the magnitudes
  of a subset of objects.  This too did not alter the basic results.

However, ubercal is only as good as its input photometry.  If the
input photometry is biased in some way, then the results will be
precisely wrong.  The most obvious kind of bias is seeing dependence.
If the photometry is not very robust against seeing changes, it may
capture a smaller fraction of the light in poor seeing conditions, and
thus ubercal will overestimate the extinction in those exposures.
After correction by ubercal, those exposures will be too bright.  We
searched for this effect in two ways.  First, we looked for
correlations between seeing and airmass terms on photometric nights,
and found no evidence for such correlations.  Second, we checked the
final photometry on the stacked images (after all corrections, as
described below) and compared the stellar locus in color-color space
between subfields which on average had good seeing in $\z$ and those
which had poor seeing in $\z$.  (Due to the observing strategy, $\z$
had the biggest seeing variations, so the other bands served as
relatively stable references for this test.)  We did not find any
indication that ubercal was fed seeing-dependent photometry.

 Because $\z$ band has the biggest seeing variations, has the
  least-clean ubercal fit, and shows the biggest variations in the
  final stacked photometry, we probed the systematics of its ubercal
  input photometry two more ways.  First, we analyzed the fit
  residuals of stars and galaxies independently. For each exposure, we
  tabulated the mean residual difference between stars and galaxies
  and performed a regression against seeing and sky noise.  There was
  a trend with seeing, such that stars were measured to be 3 mmag
  brighter in the best seeing compared to typical seeing, 3 mmag
  fainter in 10th percentile (i.e. nearly the worst) seeing, and up to
  7 mmag fainter in the very worst seeing.  The trend with sky rms was
  of a similar size but opposite sign: increased sky noise penalizes
  galaxies more than stars.  Thus the input photometry does contain
  systematic errors, but at a level lower than we are concerned with
  here.

Second, we made a completely new and independent $\z$ input catalogue
using a different algorithm, PSF fitting, which should be robust
against losing a larger fraction of object light in poor seeing.  For
each CCD of each $\z$ exposure, we compiled a list of point sources by
drawing a rectangular box around the stellar locus in size-magnitude
space.  Note that ``point source'' here refers to appearance on a
specific exposure.  If a bright star is saturated in some exposures
but not in others, it is considered a point source for these purposes
only in the latter.  We threw out observations with peak flux 80\% or
more of the already-cautious nominal saturation level.  We also
imposed a minimum peak flux of 500 ADU, for a peak S/N of about 10, to
avoid wasting time fitting low S/N stars.  We then fit elliptical
Gaussians to each point source, throwing out those which failed to
converge and those where the centroid shifted from the original
SExtractor position by more than two pixels (which could indicate a
problem such as a close neighbor).

After running ubercal on this input photometry, we found that the mean
flatfield corrections were nearly identical to those for the original
set of photometry.  More importantly, we restacked the subfield with
the most extreme seeing variations in $\z$ ($\sim$0.9-1.5\arcs\ FWHM)
and found no substantial change in the photometry relative to SDSS.
We conclude that the corrections derived here are robust against
seeing variations and robust against changes in the procedure used to
assemble the input photometry.

\section{Impact on DLS}\label{sec-impact}

Prior to running ubercal, the DLS stacks in $\z$ showed large
($\sim$20\%) spatial variations in photometry, particularly in the
fields imaged from Kitt Peak, F1 and F2.  The primary test of
photometric uniformity was the stellar locus in color-color space.
This locus changed from subfield to subfield, not just in location but
also in shape and width.  Of course, color-color plots mix errors from
different bands.  We suspected that $\z$ band had the largest
variations, but could not prove it.  Therefore, we focused on F2,
which had the largest variations and where the overlap with SDSS
allowed us to isolate the different bands.

We took stars from the SDSS database and used their colors to assign
types.  We then used the Pickles and updated
Bruzual-Persson-Gunn-Stryker\footnote{http://www.stsci.edu/hst/observatory/cdbs/bpgs.html}
stellar spectral libraries to predict DLS magnitudes using the filter,
detector, and telescope throughput curves.  Finally, we mapped the
spatial variations in calibration directly in each filter by comparing
these predictions with magnitudes measured on the DLS stacks.  This
clearly indicated that $\z$ had the largest variations, but that other
bands had substantial variations as well.

This is consistent with the flatfield corrections subsequently derived
from ubercal: The corrections for $\z$ band are about twice as large
peak-to-valley as in the other bands at Kitt Peak, and in any given
band the Cerro Tololo corrections are smaller than the Kitt Peak
corrections.  

\begin{figure}
\centerline{\includegraphics[width=3.6in]{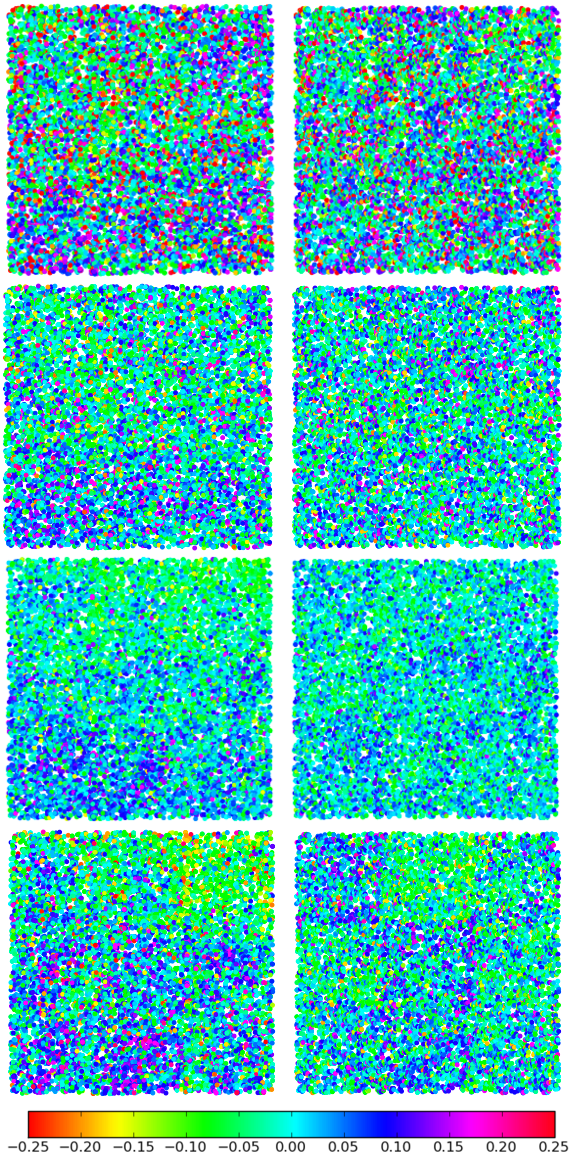}}
\caption{DLS minus SDSS magnitudes for SDSS stars in DLS field F2.
  Each image represents the 2$^\circ\times 2^\circ$ field with north
  up and east left.  The left column of images is pre-ubercal, the
  right column is post-ubercal, and the rows represent, from top to
  bottom, {\it BVRz}.  Mean values have been removed from each image
  so that nonuniformities can be more clearly compared.  Ubercal was
  successful at removing the intra-subfield bowl patterns and the
  large-scale gradients. \label{fig-sdss}}\end{figure}

Figure~\ref{fig-sdss} shows the F2 DLS-SDSS maps for DLS filters {\it
  BVRz} top to bottom, before (left) and after (right) applying
ubercal to the DLS data.  The spatial coordinates are now RA,DEC on
the sky rather than $x,y$ on the camera, and each image represents the
entire 2$^\circ\times 2^\circ$ field, which is composed of a
3$\times$3 grid of roughly camera-sized subfields.  North is up and
east is left.  There is now a lot of noise, because there are only
about 12,000 SDSS stars in this field, each star is measured only once
(on the stacked DLS image), and stars can be mistyped.  Each star is
one point on the map.  It is clear in $R$ and $\z$ at least that the
pre-ubercal stacks objects near the centers of the subfields are
fainter than objects near the edges, just as one would guess from
ubercal's flatfield corrections.  This pattern is not evident in the
post-ubercal data.  Furthermore, the pre-ubercal data have gradients
across the entire field, for example southeast to northwest in $R$ and
$\z$.  These are also not evident in the post-ubercal data.  Because
of the removal of these patterns, residual subfield-to-subfield
patterns are more evident in the post-ubercal stacks.  The continued
existence of some patterns may have more to do with the
stack-photometry issues discussed at the end of \S\ref{subsec-closeloop}
than with ubercal.

The origin of the pre-ubercal large-scale gradients may be due to a
camera-scale effect.  Referring to $\z$ band as a specific example,
imagine the north and west sides of the camera have a photometric
offset relative to the south and east sides.  The north side of the
camera in one subfield overlaps the south side of the camera in
another subfield.  When photometric offsets between subfields are
determined by enforcing consistency in the overlapping objects, this
camera-scale error gets replicated and amplified as subfields are
stitched together across a larger field.  Thus, large fields {\it
  cannot} be synthesised properly without accurately modeling
camera-scale errors.

\section{Conclusions}\label{sec-con}

We have implemented a version of the global linear least squares
procedure introduced by P08 (and known as ubercal) for the Deep Lens
Survey.  The DLS is structured very differently from SDSS, and this
required a restructuring and reimplementation of the procedure.  The
lessons learned may be of interest for surveys structured more like
the DLS, e.g., shift-and-stare rather than drift-scan.

{\it Isolated fields.}  Many deep ground-based surveys have isolated
fields which allow full nights to be split between a small number of
fields.  Our original intent was to calibrate each field against
standard stars, but with ubercal we have tied the fields together
first, and saved the overall calibration for last.  This makes the
photometry as uniform as possible.  Multiple fields can be tied
together directly only if they are observed on the same photometric
night {\it in the same filter.}  Unlike SDSS, most observers must
choose one filter at a time.  Therefore, when multiple fields are
observed on a photometric night, observers should take care to make
observations in as many relevant filters as practical.

{\it Multiple observatories.} The value of taking some observations of
southern fields from the northern telescope and vice versa was not
obvious at the start of the survey.  But in the end it had enormous
value for tying the northern and southern fields to a common system
without any reference to standard stars.  Thus, the observing plan for
photometric nights should include a small amount of time for
cross-hemisphere observing, even on fields which are not well placed
for such observing.  The point is not to go deep on those
disadvantageous fields, but just to get some cross-calibration.
Furthermore, {\it each filter} requires some cross-observing in
photometric conditions.  Because this was not in the original plan for
the DLS, we made creative use of repeated observations of equatorial
standard star fields from both observatories.  For ubercal, the value
is not in the standard stars themselves but merely in the fact that
those fields were observed by both observatories in photometric
conditions.  All objects in the fields, not just the standard stars,
were used for ubercal.

{\it Shift-and-stare imaging.} The SDSS flatfield corrections were
one-dimensional due to the drift-scan design of SDSS.  With
shift-and-stare imaging, the flatfield corrections are two-dimensional
and there are many more options for modeling them.  Although the
patterns we find are mostly smoothly varying over the focal plane, we
do find some chip-to-chip discontinuities and we believe it is
important to allow these in the model.  For Mosaic and Mosaic 2, a
second-order polynomial does reasonably well, but not perfectly.  In
\S\ref{sec-fff} we offer a suggestion for capturing high spatial
frequency features without introducing many new parameters.  

Separately, we find that the mere existence of many dithers on a given
field does {\it not} lead to good constraints on the flatfield
corrections, due to a spatiotemporal degeneracy
(\S\ref{subsec-spatiotemporal}).  There must be a strong constraint on
the time behavior of the throughput, which in practice means that
photometric nights are key to providing these constraints.

{\it Nonphotometric imaging: everything is connected.} A substantial
fraction of our imaging was taken in nonphotometric
conditions. Although photometric nights are key as explained above,
flatfield corrections can still be derived even for a completely
nonphotometric run.  This is because overlapping photometric data
provide the necessary constraints.  For example, if the true star
magnitudes in a field are well constrained by photometric nights, then
in principle one could derive flatfield corrections for any subsequent
data simply by direct comparison to the true magnitudes.  This should
work two steps removed as well: if one field is well-constrained and a
nonphotometric run includes that field as well as another field, then
the flatfield corrections can be derived by comparison to the
well-constrained field, and the flatfields then used to constrain the
spatial pattern of true magnitudes in the second field, up to a
constant offset.  Of course, a global simultaneous fit is preferred to
this multi-step procedure, but this illustrates conceptually how a
surprising amount of information can be gleaned from interlocking
observations.  In summary, ubercal performed well in solving for model
parameters on nonphotometric runs, as long as {\it some} relevant
photometric observations were available.

{\it Atmospheric variations.}  Even on photometric nights, there is a
$\sim$0.01 mag variation in atmospheric throughput on $\sim$2 hour
timescales. If there is a spatial component to this variation, future
surveys which need to be very precise might benefit from a boresighted
monitoring telescope.  Data from the monitoring telescope could
provide the more precise atmospheric model necessary to model out
these variations.

{\it Overall value of global fitting.} Applying ubercal to the DLS
revealed serious flaws in the flatfielding, and correcting these flaws
resulted in much more uniform photometry, reducing the peak-to-valley
variation from $\sim$0.2 mag to $\sim$0.05 mag, with most of the
remainder due to the non-ubercal-related factors discussed in
\S\ref{subsec-closeloop}.  Sky or dome flats are still important
because they provide pixel-by-pixel characterisation, but no observer
should rely on flats to accurately represent large-scale spatial
sensitivity variations without an independent check such as this
procedure.

Survey pipelines should always solve for these corrections to make the
photometry as uniform as possible.  In fact, without these corrections
the synthesis of large fields from many camera pointings can be very
problematic.  Because the corrections are relatively consistent from
run to run, they can also be included in quick-look reductions.  The
improvement in photometric uniformity may even be enough to influence
some observing decisions such as whether a newly discovered variable
object is interesting enough to warrant further followup.  Surveys for
which transient alerts are a data product, such as LSST, should apply
recently derived corrections in their real-time reductions to improve
the quality of the alerts.

Future surveys which wish to push the precision further may have to
consider some more subtle effects not considered here. First, by using
bright objects to derive corrections for the faint objects we are most
interested in, we have assumed some some degree of linearity. Future
surveys should control for this potential systematic error.  Second,
we have assumed that these corrections are color-independent, or at
least that correcting for the typical color of ubercal objects is
sufficient.  Future surveys will want to control for this as well, at
least by checking the color dependence if not implementing a
color-dependent corection.  The method of Stubbs \etal\ (2007) should
help greatly in controlling color dependence.

The global fitting also revealed that some exposures which were
apparently good upon manual inspection had some serious flaws.  Global
fitting thus provides a uniform, objective check on the quality of
data.  Recalling some of the more problematic data, it seems that
large surveys contain data weirder than we suppose, but not
necessarily weirder than we {\it can} suppose.  Global fitting alerted
us to the presence of unexpected effects such as apparent gain
changes, but once these effects are recognised they can be modeled
well enough to yield uniform photometry even if the ultimate physical
mechanism remains unidentified.

\section*{Acknowledgments}
We thank Nikhil Padmanabhan for helpful discussions and for providing
the basic minimisation code, Ian dell'Antonio and Robert Lupton for
helpful discussions, and Tony Tyson for making the DLS possible.
Based on observations at Kitt Peak National Observatory and Cerro
Tololo Inter-American Observatory, which are operated by the
Association of Universities for Research in Astronomy (AURA) under a
cooperative agreement with the National Science Foundation.  The Deep
Lens Survey has received major funding from Lucent Technologies and
from the National Science Foundation (grants AST-0134753, AST0441072
and AST-0708433).

{}

\end{document}